\documentclass[twocolumn,showpacs,preprintnumbers,amsmath,amssymb]{revtex4}

\usepackage{amssymb}
\usepackage{amsmath}
\usepackage{float}
\usepackage{graphicx}
\usepackage{subfigure}
\usepackage{color}
\usepackage{longtable}
\def\mk{{\mathbf p}}
\def\mq{{\mathbf q}}

\begin{document}

\title{Analyses of multi-pion Bose-Einstein correlations for granular
sources with coherent pion-emission droplets}

\author{Ghulam Bary$^1$, Wei-Ning Zhang$^{1,2}$\footnote{wnzhang@dlut.edu.cn},
Peng Ru$^3$, Jing Yang$^4$}
\affiliation{$^1$School of Physics, Dalian University of Technology, Dalian,
116024, China\\
$^2$School of Physics, Harbin Institute of Technology, Harbin, 150006, China\\
$^3$Institute of Quantum Matter, South China Normal University, Guangzhou,
510006, China\\
$^4$School of Physics, Changchun Normal University, Changchun, 130032, China}

\begin{abstract}
The ALICE Collaboration measure the three- and four-pion Bose-Einstein correlations
(BECs) in Pb-Pb collisions at the Large Hadron Collider (LHC). It is speculated
that the significant suppressions of multi-pion BECs are due to a considerable
degree of coherent pion emission in the collisions. In this paper, we study the
multi-pion BEC functions in a granular source model with coherent pion-emission
droplets. We find that the intercepts of multi-pion correlation functions at the relative momenta near zero are sensitive to droplet number in the granular source. They decrease with decreasing droplet number. The three-pion correlation
functions for the evolving granular sources with momentum-dependent partially
coherent pion-emission droplets are in basic agreement with the experimental
data in Pb-Pb collisions at $\sqrt{s_{NN}}=2.76$ TeV at the LHC. However, the model
results of four-pion correlation function are inconsistent with the experimental
data. The investigations of the normalized multi-pion correlation functions of the
granular sources indicate that there is an interesting enhancement of the normalized
four-pion correlation function in moderate relative-momentum region.
\end{abstract}
\maketitle

\section{Introduction}
Identical pion intensity correlations (Bose-Einstein correlations, BECs)
are important observables in high-energy heavy-ion collisions
\cite{Gyu79,Wongbook,Wie99,Wei00,Csorgo02,Lisa05}.
Because the multiplicity of identical pions in heavy-ion collisions at the
Large Hadron Collider (LHC) is very high, multi-pion BEC analyses with high
statistic accuracy are available \cite{ALICE-PRC14,ALICE-PRC16}.
Recently, the ALICE Collaboration measured significant suppressions of
three- and four-pion BECs in Pb-Pb collisions at $\sqrt{s_{NN}}=
2.76$ TeV at the LHC \cite{ALICE-PRC14,ALICE-PRC16}, which indicates that there
may exist a considerable coherence degree for the particle-emitting sources
produced in the collisions
\cite{ALICE-PRC14,ALICE-PRC16,Gangadharan15,BaryRuZhang18,BaryRuZhang19}.

Analyses of multi-pion BECs can provide more information of the particle-emitting
sources compared with two-pion interferometry \cite{{Wei00,Csorgo02,Liu86,Zajc87,
Biy90,And91-93,Pratt-PLB93,CsorgoZimanyi97,Zha93-00,ChaGaoZha95,HeiZhaSug,
Nak99-00,NA44,WA98,STAR-PRL03,Csa06,MorMurNak06,ALICE-PRC14,Gangadharan15,
ALICE-PRC16,BaryRuZhang18,BaryRuZhang19}}.
In particular, multi-pion BECs are sensitive to source coherence \cite{Liu86,HeiZhaSug,Gangadharan15,BaryRuZhang18,BaryRuZhang19}.
In Refs. \cite{BaryRuZhang18,BaryRuZhang19}, we investigated three- and
four-pion BECs for a spherical evolving source of pion gas with identical boson
condensation. However, the particle-emitting sources produced in relativistic
heavy-ion collisions are anisotropic in space and may have complex structure.
It is of interest to explain the experimental measurements of multi-pion BEC
suppressions at the LHC based on a more realistic model that can also explain
the other observables in the collisions.

On an event-by-event basis, the initial systems produced in relativistic heavy-ion
collisions are highly fluctuated in space. This initial fluctuation may lead to
an inhomogeneous particle-emitting source, in which there are hot spots and cold
valleys. In Refs. \cite{WNZhang04,WNZhang06,WNZhang09,WNZhang11}, a granular
source model was proposed and developed by Zhang et al. to explain the
experimental results of two-pion interferometry at the Relativistic Heavy Ion
Collider (RHIC) and LHC \cite{PHENIX-PRL04,STAR-PRC05,ALICE-PLB11}.
In Refs. \cite{YangRenZhang15,YangRenZhang15a,YangZhangRen17}, the granular
source model was used to systemically study the pion transverse-momentum spectra,
elliptic flows, and two-pion BECs in heavy-ion collisions at the RHIC and LHC.
The granular source model can reproduce the experimental data of pion
transverse-momentum spectrum, elliptic flow, and two-pion interferometry radii
\cite{YangRenZhang15,YangRenZhang15a,YangZhangRen17}.
Considering identical pions are emitted from the droplets in the granular source
model and the droplet radii are much smaller than the source size, the pion
emission from one droplet is perhaps coherent in the case of high pion event
multiplicity due to the condensation of identical bosons
\cite{WongZhang07,LiuRuZhangWong14,BaryRuZhang18,BaryRuZhang19}.

In this work, we consider a granular source with coherent pion-emission droplets.
The droplets in the granular source move with anisotropic velocities and evolve
in viscous hydrodynamics, as described in Ref. \cite{YangZhangRen17}. However,
identical pion emissions from one droplet are assumed to be completely or partially
coherent. We investigate multi-pion BECs in the granular source model with
coherent pion-emission droplets. The normalized three- and four-pion correlation
functions of the granular sources are examined for completely coherent and
momentum-dependent partially coherent pion emissions from a droplet.

The rest of this paper is organized as follows. In Sec. II, we examine the
three- and four-pion BEC functions of a static granular source with coherent
pion-emission droplets. In Sec. III, we investigate the three- and four-pion
BECs in the granular source model in which the droplets evolve in viscous
hydrodynamics. We also investigate the normalized multi-pion correlation functions
of the evolving granular sources in this section. Finally, we give a summary and
discussion in Sec. IV.

\section{Multi-pion BECs of static granular sources}
We first consider a static granular source in which identical pions are emitted
from separated droplets. The spatial distribution of the emitting points in
each droplet is assumed with a Gaussian
distribution $\sim\!e^{r^2/(2r_d^2)}$, and the droplet centers $R_j~(j=1,2,
\cdots,n)$ distribute in the granular source with a Gaussian distribution $\sim\!e^{-R_j^2/2R_G^2}$. The two- and three-pion BEC functions of the
static granular source can be expressed as \cite{PraSieVis92,Zhang95}
\begin{widetext}
\begin{eqnarray}
&&C_2(\mk_1,\mk_2)=1+\frac{1}{n}e^{-q_{12}^2 r_d^2}+\bigg(1-\frac{1}{n}\bigg)
e^{-q_{12}^2 (r_d^2+R_G^2)} \equiv 1+\frac{1}{n} {\cal R}^d(1,2)+\bigg(1
-\frac{1}{n}\bigg){\cal R}^G(1,2), ~~~~
\label{GsC2}\\
&&C_3(\mk_1,\mk_2,\mk_3)=1+\frac{1}{n}\bigg[{\cal R}^d(1,2)+{\cal R}^d(1,3)
+{\cal R}^d(2,3)\bigg]+\bigg(1-\frac{1}{n}\bigg)\bigg[{\cal R}^G(1,2)
+{\cal R}^G(1,3)+{\cal R}^G(2,3)\bigg]~~~~~~~~~~~~~\cr
&&\hspace*{29mm}+\frac{2}{n^2}\bigg[{\cal R}^d(1,2){\cal R}^d(1,3){\cal R}^d(2,3)
\bigg]^{\frac{1}{2}}+\frac{2(n-1)}{n^2}\bigg[\Big({\cal R}^d(1,3){\cal R}^d(2,3)/
{\cal R}^d(1,2)\Big)^{\!\frac{1}{2}}{\cal R}^G(1,2)\cr
&&\hspace*{29mm}+\Big({\cal R}^d(1,2){\cal R}^d(2,3)/{\cal R}^d(1,3)\Big)^{\!\frac{1}{2}}{\cal R}^G(1,3)+\Big({\cal R}^d(1,2){\cal R}^d(1,3)/
{\cal R}^d(2,3)\Big)^{\!\frac{1}{2}}{\cal R}^G(2,3)\bigg]\cr
&&\hspace*{29mm}+\frac{2(n-1)(n-2)}{n^2}\bigg[{\cal R}^G(1,2){\cal R}^G(1,3)
{\cal R}^G(2,3)\bigg]^{\frac{1}{2}},
\label{GsC3}
\end{eqnarray}
\end{widetext}
where $n$ is the droplet number in the granular source,
${\cal R}^d(1,2)=e^{-\mq_{12}^2 r_d^2}$, ${\cal R}^G(1, 2)=e^{-\mq_{12}^2 (r_d^2+R_G^2)}$,and $\mq_{ij}=\mk_i-\mk_j~(i,j=1,2,3,4)$.
In Eq.~(\ref{GsC2}), the ${\cal R}^d(1,2)$ and ${\cal R}^G(1,2)$ terms express
the correlations of two pions emitted from one droplet and different droplets,
respectively. In Eq.~(\ref{GsC3}), the $\frac{2}{n^2}[\,\cdots\,]$ and
$\frac{2(n-1)}{n^2}[\,\cdots\,]$ terms express the pure triplet correlations of
three pions emitted from one droplet and two pions emitted from one droplet,
respectively. In Eq.~(\ref{GsC3}), the last term expresses the pure triplet
correlations of three pions emitted from different droplets. The two- and
three-pion correlation functions become those of the Gaussian sources when
$n\to\infty~(r_d\to 0)$. Similarly, one can get the four-pion correlation
function of the static granular source as given in Appendix A.

For a small droplet radius, the pion emission from a droplet is significantly
coherent \cite{WongZhang07,LiuRuZhangWong14,BaryRuZhang18,BaryRuZhang19}.
Assuming the pions emitted from one droplet are completely coherent, the
two- and three-pion correlation functions of the granular source become
\begin{equation}
C_2(\mk_1,\mk_2)=1+\frac{(n-1)}{n}{\cal R}^G(1,2),
\label{GsCC2}
\end{equation}
\begin{widetext}
\begin{eqnarray}
C_3(\mk_1,\mk_2,\mk_3)\!=\!1\!+\!\frac{(n-1)}{n}\bigg[{\cal R}^G(1,2)
\!+\!{\cal R}^G(1,3)\!+\!{\cal R}^G(2,3)\bigg] + \frac{2(n-1)(n-2)}{n^2}
\bigg[{\cal R}^G(1,2){\cal R}^G(1,3){\cal R}^G(2,3)\bigg]^{\!\frac{1}{2}}.
\label{GsCC3}
\end{eqnarray}
The four-pion correlation function of the granular source with completely
coherent pion-emission droplets can be expressed as

\begin{figure}
\centering
\includegraphics[width=0.82\columnwidth]{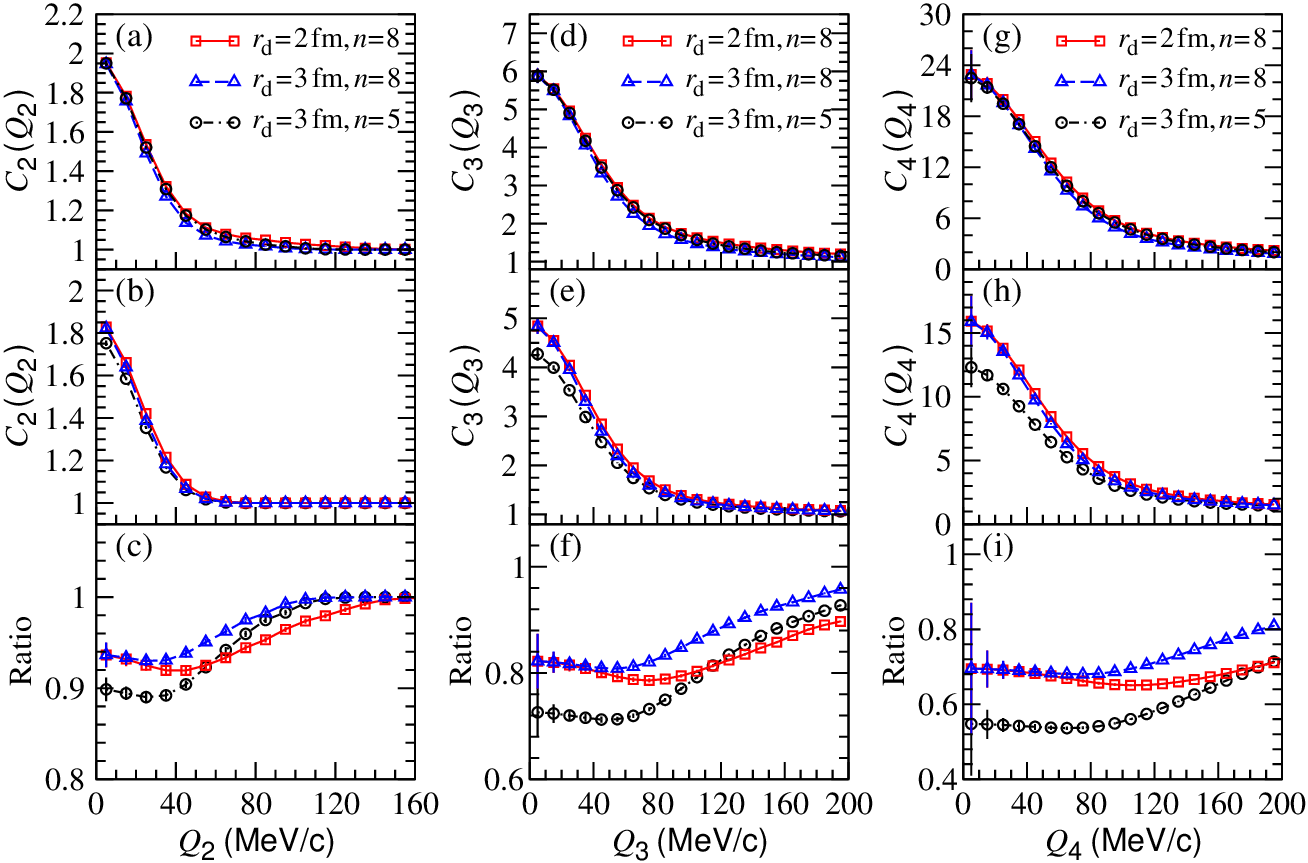}
\caption{(Color online) Two-, three- and four-pion correlation functions of the
static granular source with chaotic (top panels) and completely coherent (middle
panels) pion-emission droplets. Here, the radii of the granular sources are taken
to be $R_G=6.0$ fm. The bottom panels are the ratios of the correlation functions
of the granular sources with the completely coherent droplets to the granular
sources with the chaotic droplets. }
\label{SGs-C2C3}
\end{figure}

\begin{eqnarray}
&&C_4(\mk_1,\mk_2,\mk_3,\mk_4)\!=\!1+\frac{(n-1)}{n}\bigg[{\cal R}^G(1,2)
\!+\!{\cal R}^G(1,3)\!+\!{\cal R}^G(1,4)\!+\!{\cal R}^G(2,3)\!+\!{\cal R}^G
(2,4)\!+\!{\cal R}^G(3,4)\bigg]~~~~~~~~~~~~~~~~~~~~\cr
&&\hspace*{33mm}+\frac{2(n-1)(n-2)}{n^2}\bigg[\bigg({\cal R}^G(1,2){\cal R}^G(1,3)
{\cal R}^G(2,3)\bigg)^{\!\frac{1}{2}}\!+\!\bigg({\cal R}^G(1,2){\cal R}^G(1,4)
{\cal R}^G(2,4)\bigg)^{\!\frac{1}{2}}\cr
&&\hspace*{33mm}+\bigg({\cal R}^G(2,3){\cal R}^G(2,4){\cal R}^G(3,4)
\bigg)^{\!\frac{1}{2}}\!+\!\bigg({\cal R}^G(1,3){\cal R}^G(1,4)
{\cal R}^G(3,4)\bigg)^{\!\frac{1}{2}}\bigg]\cr
&&\hspace*{33mm}+\frac{(n\!-\!1)(n\!-\!2)(n\!-\!3)}{n^3}\bigg[{\cal R}^G(1,2)
{\cal R}^G(3,4)\!+\!{\cal R}^G(1,3){\cal R}^G(2,4)\!+\!{\cal R}^G(2,3){\cal R}^G(1,4)
\bigg]\cr
&&\hspace*{33mm}
+\frac{(n-1)}{n^3}\bigg[{\cal R}^G(1,2){\cal R}^G(3,4)\bigg(e^{-2\mq_{12}\cdot\mq_{34}
R_G^2} +e^{-2\mq_{12}\cdot\mq_{43}R_G^2}\bigg) +{\cal R}^G(1,3){\cal R}^G(2,4)
\cr
&&\hspace*{33mm}
\times \bigg(e^{-2\mq_{13}\cdot\mq_{24}R_G^2} +e^{-2\mq_{13}\cdot\mq_{42}R_G^2}\bigg)
+{\cal R}^G(1,4){\cal R}^G(2,3)\bigg(e^{-2\mq_{14}\cdot\mq_{23}R_G^2} +e^{-2\mq_{14}\cdot\mq_{32}R_G^2}\bigg)\bigg]
\cr
&&\hspace*{33mm}
+\frac{2(n-1)(n-2)}{n^3}\bigg[{\cal R}^G(1,2){\cal R}^G(3,4)\bigg(
e^{-\mq_{12}\cdot\mq_{34}R_G^2} +e^{-\mq_{12}\cdot\mq_{43}R_G^2}\bigg) +{\cal
R}^G(1,3){\cal R}^G(2,4)
\cr
&&\hspace*{33mm}
\times \bigg(e^{-\mq_{13}\cdot\mq_{24}R_G^2} +e^{-\mq_{13}\cdot\mq_{42}R_G^2}\bigg)
+{\cal R}^G(1,4){\cal R}^G(2,3)\bigg(e^{-\mq_{14}\cdot\mq_{23}R_G^2} +e^{-\mq_{14}\cdot\mq_{32}R_G^2}\bigg)\bigg]
\cr
&&\hspace*{33mm}+\frac{2(n-1)(n-2)(n-3)}{n^3}\bigg[\bigg({\cal R}^G(1,2){\cal R}^G(2,3){\cal R}^G(3,4){\cal R}^G(1,4)\bigg)^{\!\frac{1}{2}}\cr
&&\hspace*{33mm}+\bigg({\cal R}^G(1,3){\cal R}^G(2,3){\cal R}^G(2,4){\cal R}^G(1,4)\bigg)^{\!\frac{1}{2}}\!+\!\bigg({\cal R}^G(1,2){\cal R}^G(2,4)
{\cal R}^G(3,4){\cal R}^G(1,3)\bigg)^{\!\frac{1}{2}}\bigg],
\label{GsC4}
\end{eqnarray}
\end{widetext}
where the third square brackets are the correlations of double pion pairs where four
pions are emitted from four different droplets; the fourth square brackets are the
correlations of double pion pairs where the two pions of a pair are emitted from two
different droplets and the two pions of another pair are emitted respectively from
the two droplets also; the fifth square brackets are the correlations of double pion
pairs where the two pions of a pair are emitted from two different droplets and the
two pions of another pair are emitted respectively from one of the same droplets and
from another droplet; and the last square brackets are the pure quadruplet pion
correlations where four pions emitted from different four droplets. A detailed
deviation of the correlation function may see in Appendix A.

In Figs.~\ref{SGs-C2C3}(a) and \ref{SGs-C2C3}(b), we plot the two-pion correlation
functions of the static granular sources with chaotic and completely coherent
pion-emission droplets, respectively. In Figs.~\ref{SGs-C2C3}(d) and
\ref{SGs-C2C3}(e), we plot the three-pion correlation functions of the static
granular sources with chaotic and completely coherent pion-emission droplets,
respectively. In Figs.~\ref{SGs-C2C3}(g) and \ref{SGs-C2C3}(h), we plot the four-pion
correlation functions of the static granular sources with chaotic and completely
coherent pion-emission droplets, respectively. The panels (c), (f) and (i) show the
ratios of the correlation functions of the granular sources with completely coherent
pion-emission droplets to the correlation functions of the granular source with
chaotic pion-emission droplets. Here, the radii of the granular sources are taken
to be $R_G=6.0$ fm.
The variable $Q_m~(m=2,3,\cdots)$ is defined by the covariant relative momenta, as
\begin{equation}
Q_m=\sqrt{\sum_{i<j\le m}-(p_i-p_j)^\mu (p_i-p_j)_\mu}, ~~~~(m\ge2).
\end{equation}

From Fig.~\ref{SGs-C2C3} it can be seen that the intercepts of the correlation
functions of the granular sources with the completely coherent droplets decrease obviously compared with the granular sources with the chaotic droplets.
The intercept decreases will become small with increasing droplet number $n$. The
intercept of the four-pion correlation function decreases much than those of two-
and three-pion correlation functions for a fixed droplet number $n$. From the ratio
results one can see that the correlation functions for the the completely coherent
droplets have the more decreases for a small droplet radius in the high relative-momentum variable regions.

\begin{figure}[tbp]
\centering
\includegraphics[width=0.75\columnwidth]{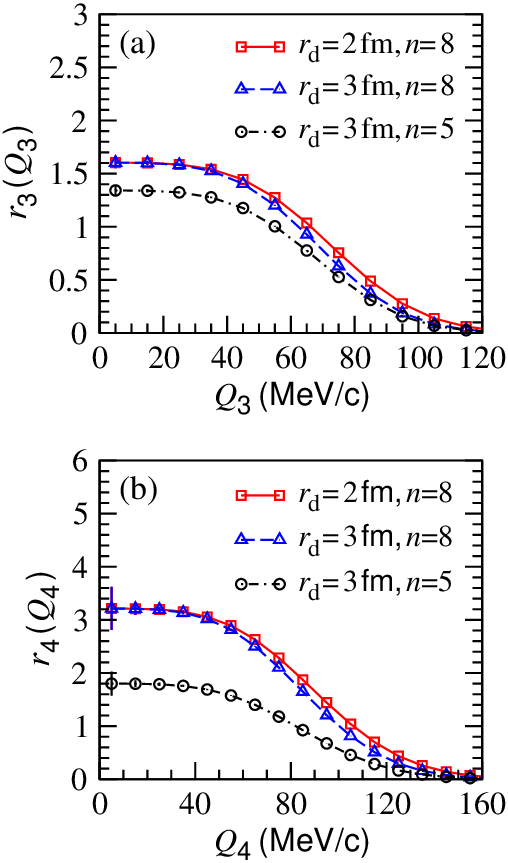}
\caption{(Color online) (a) Normalized three-pion correlation functions of the
static granular sources with completely coherent pion-emission droplets.
(b) Normalized four-pion correlation functions of the static granular sources
with completely coherent pion-emission droplets. Parameters of the granular
sources are the same as in Fig.~\ref{SGs-C2C3}. }
\label{SGs-r3r4}
\end{figure}

\begin{figure}[tbp]
\centering
\includegraphics[width=0.68\columnwidth]{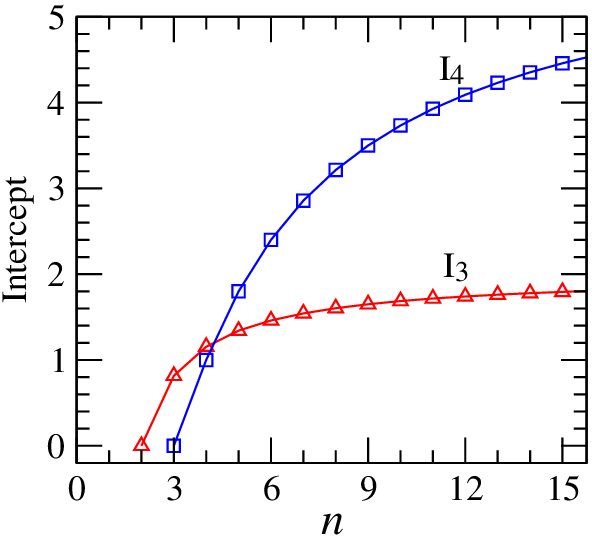}
\caption{(Color online) Intercepts of the $r_3(Q_3)$ and $r_4(Q_4)$ of
granular sources with completely coherent pion-emission droplets, as functions
of droplet number in static granular sources. }
\label{I34-n}
\end{figure}

The normalized three-pion correlation function $r_3$ is defined by the ratio
of the three-pion cumulant correlator to the square root of the product of the
two-particle correlators \cite{HeiZhaSug}. For the granular source with
completely coherent pion-emission droplets, it is given by
\begin{equation}
\label{smallr3}
r_3(Q_3)=\frac{[c_3(Q_3)-1][n/(n-1)]^{3/2}}
{\sqrt{{\cal R}^G(1,2)\!(Q_3){\cal R}^G(2,3)\!(Q_3){\cal R}^G(1,3)\!(Q_3)}},
\end{equation}
where
\begin{eqnarray}
&&c_3(Q_3)=1+\frac{2(n-1)(n-2)}{n^2}\cr
&&\hspace*{15mm}\times \bigg[{\cal R}^G(1,2){\cal R}^G(1,3){\cal R}^G(2,3)
\bigg]^{\!\frac{1}{2}}\!\!\!(Q_3).~~~~~~~
\label{Gsc3}
\end{eqnarray}
Because $r_3$ is insensitive to resonance decay, it is used to measure source
coherence in analyses of experimental data \cite{NA44,WA98,STAR-PRL03,ALICE-PRC14}.
Similarly, the normalized four-pion correlation function $r_4$ of the granular
source is given by
\begin{eqnarray}
\label{smallr4}
&&\hspace*{-4mm}r_4(Q_4)=\cr
&&\hspace*{-2mm}\frac{[c_4(Q_4)-1][n/(n-1)]^2}{\sqrt{{\cal R}^G(1,2)\!(Q_4){\cal
R}^G(2,3)\!(Q_4) {\cal R}^G(3,4)\!(Q_4){\cal R}^G(1,4)\!(Q_4)}}, \nonumber\\
\end{eqnarray}
where
\begin{eqnarray}
&&c_4(Q_4)=1+\frac{2(n-1)(n-2)(n-3)}{n^3}\cr
&&\hspace*{0mm}\times\bigg[\bigg({\cal R}^G(1,2){\cal R}^G(2,3)
{\cal R}^G(3,4){\cal R}^G(1,4)\bigg)^{\!\frac{1}{2}}\!\!\!(Q_4)\cr
&&\hspace*{2mm}+\bigg({\cal R}^G(1,3){\cal R}^G(2,3){\cal R}^G(2,4){\cal R}^G(1,4)\bigg)^{\!\frac{1}{2}}\!\!\!(Q_4)\cr
&&\hspace*{2mm}+\bigg({\cal R}^G(1,2){\cal R}^G(2,4){\cal R}^G(3,4)
{\cal R}^G(1,3)\bigg)^{\!\frac{1}{2}}\!\!\!(Q_4)\bigg].~~~~
\end{eqnarray}

In Figs. \ref{SGs-r3r4}(a) and \ref{SGs-r3r4}(b), we plot the normalized
three- and four-pion correlation functions $r_3(Q_3)$ and $r_4(Q_4)$ of the
static granular sources with completely coherent pion-emission droplets,
respectively. The parameters of the granular source are the same as in
Fig.~\ref{SGs-C2C3}. The normalized correlation functions have
plateaus in low-$Q_{3,4}$ regions to $\sim\!50$ MeV$\!/c$.
We plot the intercepts of the $r_3(Q_3)$ and $r_4(Q_4)$ as functions of
droplet number in Fig.~\ref{I34-n}. The intercepts of four-pion normalized
correlation functions are more sensitive to the droplet number $n$ than
three-pion normalized correlation functions in the $5<n<12$ region.

Because without expansion, the three- and four-pion correlation functions of static
granular sources fall rapidly with the multi-pion relative momenta $Q_3$ and $Q_4$,
respectively. It is hard to describe the experimental data of multi-pion correlations
with a static granular source model. We shall investigate in next section the three-
and four-pion correlation functions in an evolving granular source model and compare
the multi-pion correlation functions with experimental data.

\section{Multi-pion BECs in evolving granular source model}
The evolving granular source models can reproduce the pion transverse-momentum
spectrum, elliptic flow, and interferometry radii
\cite{YangRenZhang15,YangRenZhang15a,YangZhangRen17}.
We next investigate the multi-pion BECs for the evolving granular sources in
which the droplets expand in viscous hydrodynamics and emit pions coherently.

\subsection{Evolving granular source model}
The model we consider is based on a viscous granular source model developed in
Ref.~\cite{YangZhangRen17}, but here the pions emitted from one droplet are
assumed to be completely or partially coherent. In this subsection, we briefly present
the ingredients of the granular source model used in the work. For details of
granular source models, the reader is referred to Refs.
\cite{WNZhang06,WNZhang09,WNZhang11,YangRenZhang15,YangRenZhang15a,YangZhangRen17}.

The granular source model was proposed by W. N. Zhang {\it et al.}
\cite{WNZhang04,WNZhang06}, to explain the RHIC HBT puzzle,
$R_{\text{out}}/R_{\text{side}}~\sim~1$
\cite{STAR-PRL01,PHENIX-PRL02,PHENIX-PRL04,STAR-PRC05},
where $R_{\text{out}}$ and $R_{\text{side}}$ are two HBT radii in transverse
plane along and perpendicular to the transverse momentum of the particle
pairs \cite{Bertsch88,Pratt90}. Because the HBT radius $R_{\text{out}}$ decrease
with decreasing system lifetime and the HBT $R_{\text{side}}$ increase with
increasing system size, a particle-emitting source with small hot droplets,
therefore small evolution time, and distributed in a large space range may
lead to a result of $R_{\text{out}}/R_{\text{side}}~\sim~1$
\cite{WNZhang04,WNZhang06,WNZhang09,WNZhang11}.
Although the early idea of constructing a granular source model was based on the
first-order QCD transition, the occurrence of granular sources may not be limited to
first-order phase transition. In relativistic heavy ion collisions at the RHIC and LHC
energies, the large system initial fluctuations and some instabilities in the early
violent expansion of the system may lead to the granular inhomogeneous structure of
the particle-emitting sources \cite{WNZhang06,WNZhang09,WNZhang11}.

In the granular source model, it is assumed that the initial spatial inhomogeneous
and violent expansion at the early stages of the system produced in ultrarelativistic
heavy-ion collisions may lead to a breakup of the system to many hot and dense droplets
and formation of a granular particle-emitting source. At the formation time of the initial granular source, the droplet centers distribute within a cylinder along the
collision axis, and the initial energy distribution in a droplet satisfies a
Woods-Saxon distribution as in Refs. \cite{YangRenZhang15a,YangZhangRen17}.
The average droplet number, $\langle n\rangle$, of the granular source is related
to the initial mean separation and geometry of the source \cite{PraSieVis92}.

The evolution of the granular source includes the droplet evolution in viscous
hydrodynamics and the droplet expansion in whole with anisotropic droplet
velocities $v_{dx}$, $v_{dy}$, and $v_{dz}$. In the granular source model, the
geometry and velocity parameters are determined by comparing the model results
of pion transverse-momentum spectrum, elliptic flow, and two-pion interferometry
radii with experimental data. In this paper, we use the viscous granular source
model developed in Ref. \cite{YangZhangRen17} to describe the source evolution
in Pb-Pb collisions at $\sqrt{s_{_{NN}}}=2.76$ TeV \cite{ALICE-PRC16}, and the
model parameters are taken as the same in \cite{YangZhangRen17}.

\subsection{Multi-pion correlation functions}

\begin{figure}[tbp]
\centering
\includegraphics[width=0.88\columnwidth]{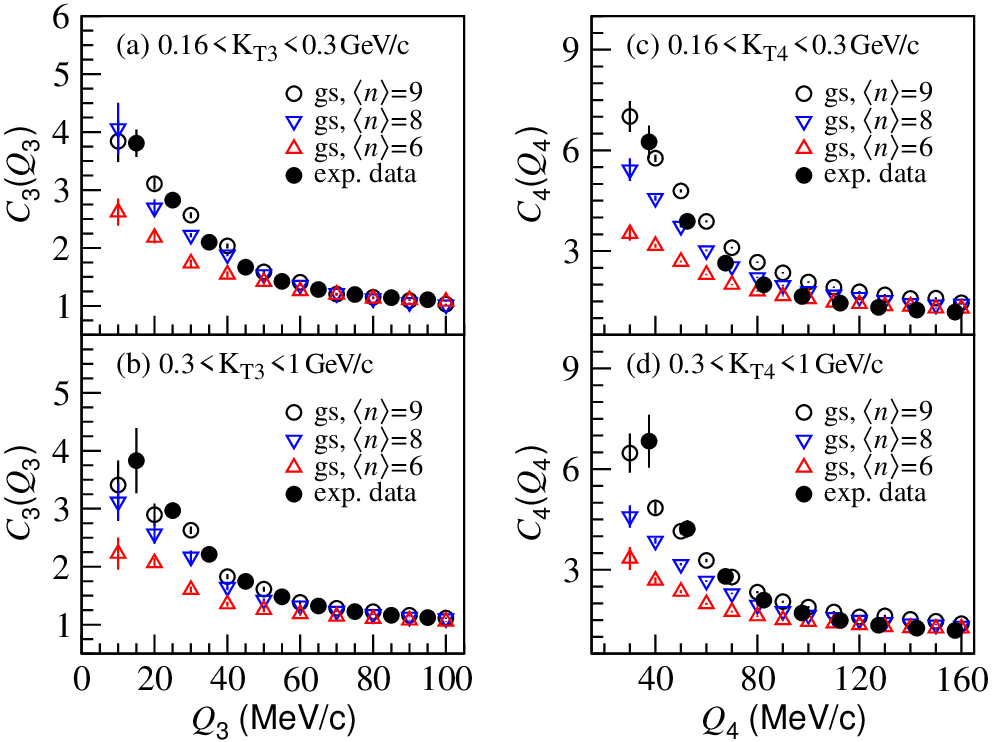}
\caption{(Color online) Three- and four-pion correlation functions of
evolving granular sources with the completely coherent droplets and
experimental data in central Pb-Pb collisions at $\sqrt{s_{_{NN}}}=2.76$
TeV \cite{ALICE-PRC16}, in transverse-momentum intervals $0.16<K_{T3}<0.3$
GeV$\!/c$ and $0.3<K_{T3}<1$ GeV$\!/c$. Here, $\langle n\rangle$ denotes
average droplet number in evolving granular sources. }
\label{EGs-C3}
\end{figure}

In Figs. \ref{EGs-C3}(a) and \ref{EGs-C3}(b), we plot the three-pion correlation
functions $C_3(Q_3)$ of the evolving granular sources with completely coherent
pion-emission droplets for central Pb-Pb collisions at $\sqrt{s_{_{NN}}}=2.76$ TeV.
The experimental data of $C_3(Q_3)$ measured by the ALICE Collaboration in central
Pb-Pb collisions \cite{ALICE-PRC16} are presented for comparison. Panels (a) and
(b) show the results in the low- and high-transverse-momentum intervals $0.16<K_{T3}
<0.3$~GeV$\!/c$ and $0.3<K_{T3}<1$ GeV$\!/c$, respectively. Here, $K_{T3}=|\mk_{T1}
+\mk_{T2}+\mk_{T3}|/3$. The average droplet number $\langle n\rangle$ for the
simulation events with the granular source parameters determined together by
the experimental data of transverse-momentum spectra, elliptic flow, and
interferometry radii in 0\%--10\% Pb-Pb collisions \cite{YangZhangRen17} is
8. In Fig. \ref{EGs-C3}, the results for $\langle n\rangle=9$ and $\langle n
\rangle=6$ are calculated with the same granular source parameters but a smaller
and larger initial mean separation. The granular source results in the small-$Q_3$
region increase with increasing average droplet number $\langle n\rangle$.
However, the results for $\langle n\rangle=8$ are lower than the experimental
data in the small-$Q_3$ region.

In Fig.~\ref{EGs-C3}(c) and \ref{EGs-C3}(d), we plot the four-pion correlation
functions $C_4(Q_4)$ of the evolving granular sources with completely coherent pion-emission droplets for central Pb-Pb collisions at $\sqrt{s_{_{NN}}}=2.76$ TeV.
The experimental data of $C_4(Q_4)$ measured by the ALICE Collaboration in central
Pb-Pb collisions \cite{ALICE-PRC16} are presented for comparison. Panels (a) and
(b) show the results in the low and high transverse-momentum intervals $0.16<K_{T4}
<0.3$ GeV$\!/c$ and $0.3<K_{T4}<1$ GeV$\!/c$, respectively. Here, $K_{T4}=|\mk_{T1}
+\mk_{T2}+\mk_{T3}+\mk_{T4}|/4$. The four-pion correlation functions of the
granular sources increase with increasing average droplet number $\langle n
\rangle$ in the small-$Q_4$ region. In addition, the results for $\langle n\rangle=8$
are lower than the experimental data in the small-$Q_4$ region.
The multi-pion correlation functions are sensitive to the droplet number in the
granular source. However, the transverse-momentum spectrum and elliptic flow of the
granular sources are insensitive to $\langle n\rangle$
\cite{YangRenZhang15,YangRenZhang15a,YangZhangRen17}.

Considering that the pions with high momenta are more possibly emitted chaotically
from the excited states \cite{WongZhang07,LiuRuZhangWong14,BaryRuZhang18,BaryRuZhang19},
we further investigate the multi-pion BECs for the granular sources with partially
coherent pion-emission droplets. We assume that the pions emitted from one droplet
and with momenta lower than a fixed value $p'$ are amplitude coherent and therefore
without intensity correlation. However, the pions with momenta higher than $p'$ are chaotic emission (from excited-states).

\begin{figure}[tbp]
\centering
\includegraphics[width=0.88\columnwidth]{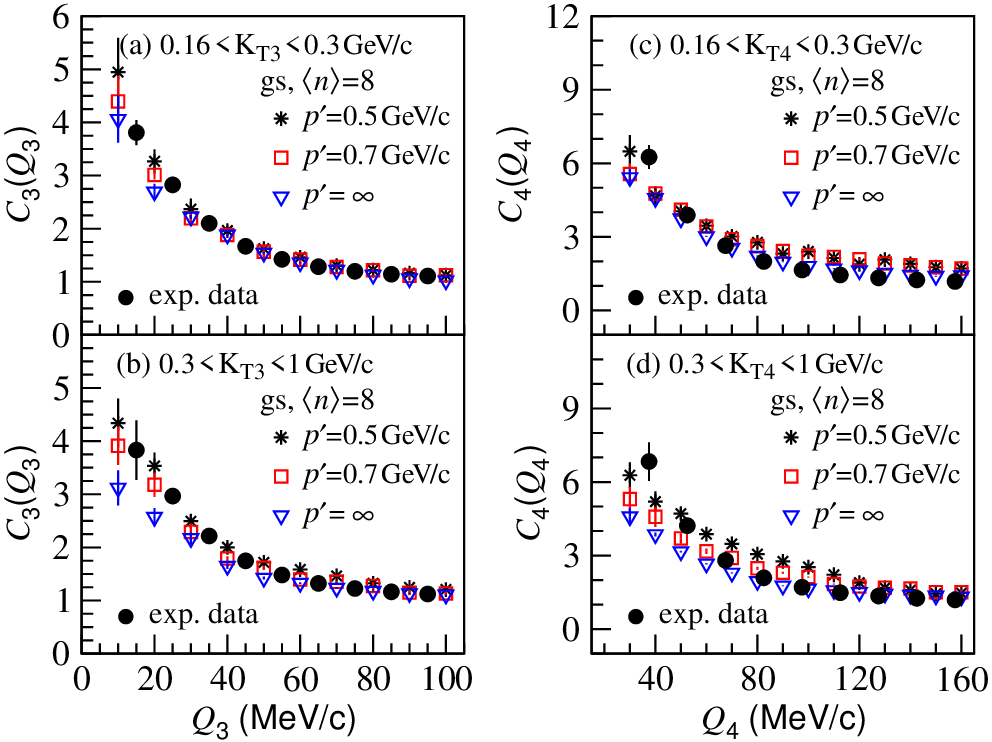}
\caption{(Color online) Three- and four-pion correlation functions of evolving
granular sources with completely coherent ($p'=\infty$) and partially coherent
($p'=$ 0.5 and 0.7 GeV$\!/c$) pion-emission droplets, in transverse-momentum
intervals $0.16<K_{T3}<0.3$ GeV$\!/c$ and $0.3<K_{T3}<1$ GeV$\!/c$. Here, average
droplet number of the granular source is 8 and solid-circle symbols are
experimental data in central Pb-Pb collisions at $\sqrt{s_{_{NN}}}=2.76$ TeV
\cite{ALICE-PRC16}. }
\label{EGs-C3-pp}
\end{figure}

In Figs. \ref{EGs-C3-pp}(a) and \ref{EGs-C3-pp}(b), we compare the three-pion
correlation functions of the evolving granular sources with completely coherent
pion-emission droplets (corresponding to $p'=\infty$) and partially coherent
pion-emission droplets with $p'=$ 0.5 and 0.7 GeV$\!/c$, in the lower- and higher-
transverse-momentum $K_{T3}$ intervals, respectively. Here, the average droplet
number of the granular sources is 8 and the solid-circle symbols are the
experimental data in central Pb-Pb collisions at $\sqrt{s_{_{NN}}}=2.76$ TeV
\cite{ALICE-PRC16}. In the low-$Q_3$ region, the three-pion correlation functions
exhibit an increase with decreasing $p'$ and the increase is greater in the higher
$K_{T3}$ interval than in the lower $K_{T3}$ interval. This is because the
contribution of chaotic pion emission in the correlation functions increases
with decreasing $p'$ and the pions with high momenta, which are more possibly
emitted chaotically, have higher $K_{T3}$ than those with low momenta. The
three-pion correlation functions of the granular source with $p'=0.5$ GeV$\!/c$
are approximately in agreement with the experimental data.

In Figs. \ref{EGs-C3-pp}(c) and \ref{EGs-C3-pp}(d), we compare the four-pion
correlation functions of the evolving granular sources with completely coherent
pion-emission droplets (corresponding to $p'=\infty$) and partially coherent
pion-emission droplets with $p'=$ 0.5 and 0.7 GeV$\!/c$, in the lower-and
higher-transverse-momentum $K_{T4}$ intervals, respectively. Here, the average droplet
number of the granular sources is 8 and the solid-circle symbols are the
experimental data in central Pb-Pb collisions at $\sqrt{s_{_{NN}}}=2.76$ TeV
\cite{ALICE-PRC16}. Compared with the three-pion correlation functions, the four-pion correlation functions of the granular source are more sensitive to the value of $p'$,
but inconsistent with the experimental data. This puzzle in the current framework
indicates that more considerations of the partially coherent pion-emission are needed
to let the model multi-pion correlation functions agree with experimental data.

\begin{figure*}[tbp]
\centering
\includegraphics[width=1.9\columnwidth]{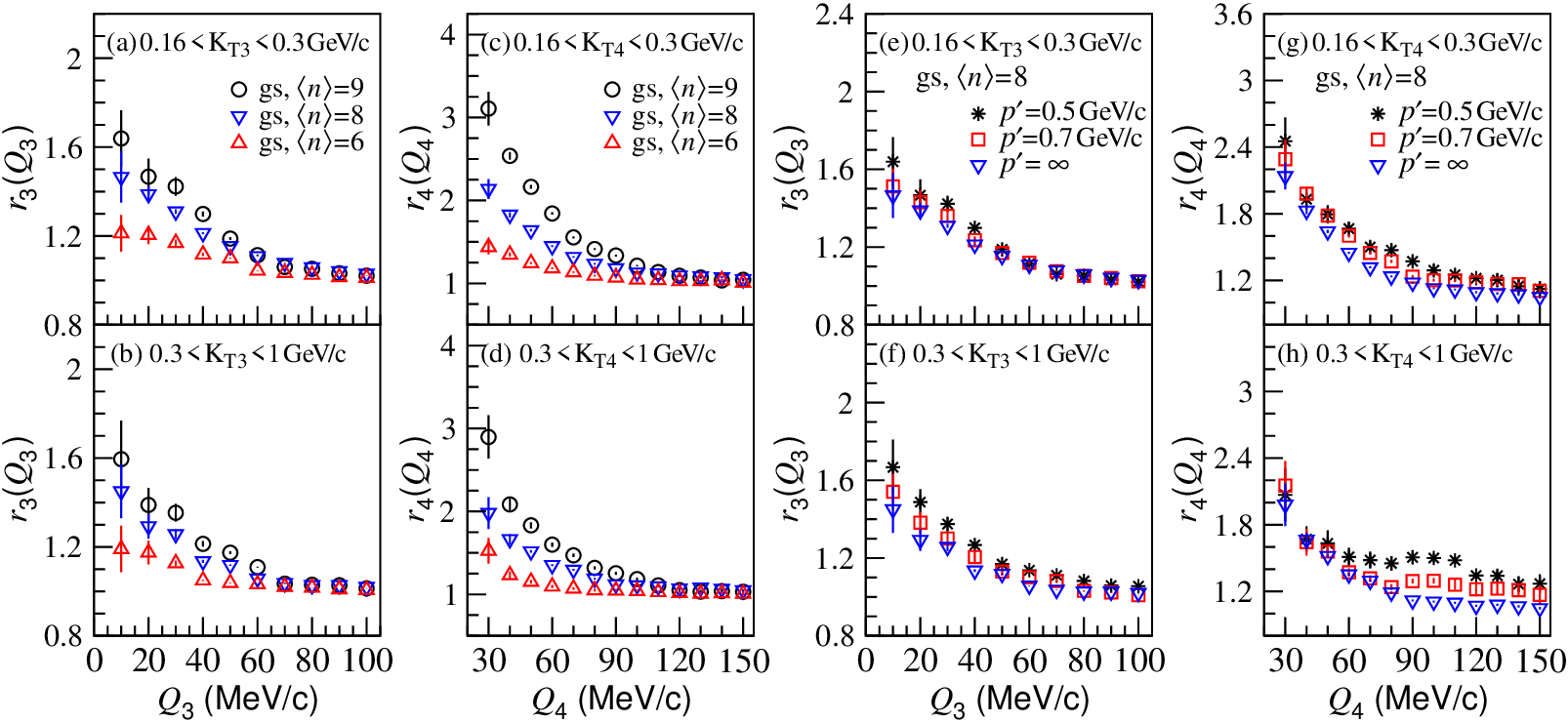}
\caption{(Color online) (a) -- (d) Normalized three- and four-pion correlation
functions of evolving granular sources with completely coherent pion-emission
droplets. (e) -- (h) Normalized three and four-pion correlation functions of
evolving granular sources with partially coherent ($p'=$ 0.5 and 0.7 GeV$\!/c$)
and completely coherent ($p'=\infty$) pion-emission droplets for the average
droplet number $\langle n\rangle=8$. }
\label{EGs-r3}
\end{figure*}

\subsection{Normalized multi-pion BEC functions}
The normalized multi-pion correlation functions $r_3$ and $r_4$ are believed to be
suitable for analyzing the source coherence in relativistic heavy-ion collisions.
In the preceding section, it was found that the normalized correlation functions are sensitive to droplet number in the static granular source with completely coherent
pion-emission droplets. In this subsection, we investigate $r_3$ and $r_4$in the
evolving granular source with completely coherent and partially coherent
pion-emission droplets.

We show in Figs.~\ref{EGs-r3}(a) and \ref{EGs-r3}(b) the normalized three-pion
correlation functions of the evolving granular sources with completely coherent
pion-emission droplets and the different values of average droplet number $\langle
n\rangle$ in the transverse-momentum intervals $0.16<K_{T3}<0.3$ GeV$\!/c$ and
$0.3<K_{T3}<1$ GeV$\!/c$, respectively. The normalized correlation function
increases with increasing $\langle n\rangle$ in both transverse-momentum
intervals. Compared with the normalized three-pion correlation functions of the
static granular sources shown in Fig. \ref{SGs-r3r4}(a), which have a
plateau structure in the small-$Q_3$ region, the results of the evolving granular
sources with the large $\langle n\rangle$ values decrease with $Q_3$ in the
small-$Q_3$ region. The decrease of $r_3$ with increasing $Q_3$ indicates that the
three-pion cumulant correlator (correlation of pure pion-triplet interference)
decreases more rapidly with increasing $Q_3$ than two-pion correlations.

We show in Figs.~\ref{EGs-r3}(c) and \ref{EGs-r3}(d) the normalized four-pion
correlation functions of the evolving granular sources with completely coherent
pion-emission droplets and the different values of average droplet number $\langle
n\rangle$ in the transverse-momentum intervals $0.16<K_{T4}<0.3$ GeV$\!/c$ and $0.3<K_{T4}<1$ GeV$\!/c$, respectively.
The normalized correlation function increases with increasing $\langle n\rangle$
in both transverse-momentum intervals. Compared with the normalized three-pion
correlation functions of the evolving granular sources,
the four-pion correlation functions are more sensitive to $\langle n\rangle$.

In Figs.~\ref{EGs-r3}(e)~--~\ref{EGs-r3}(h), we compare the normalized three- and
four-pion correlation functions of the evolving granular sources with completely
coherent ($p'=\infty$) and partially coherent ($p'=$ 0.5 and 0.7 GeV$\!/c$)
pion-emission droplets. Here, the average droplet number in the granular source is
8. The normalized three-pion correlation functions increase slightly with decreasing
$p'$. However, the intercepts of the correlation functions at $Q_3\sim 0$ are
approximately in agreement, because the intercept is mainly determined by the
droplet number in the granular source.
The normalized four-pion correlation functions increase with decreasing
$p'$. In the wide transverse-momentum interval, $0.3<K_{T4}<1$ GeV$\!/c$, the
normalized four-pion correlation function for the smallest $p'$ has an obvious
enhancement at approximately $Q_4\sim 100$ MeV$\!/c$ due to the momentum dependence
of pion-emission coherence and the sensitivity of high-order pion correlations to
source coherence. As discussed in Ref. \cite{BaryRuZhang19}, the average pion
momentum will increases with increasing $Q_4$ if there are no other constraints.
This leads to an increase of chaotic emission possibility with increasing $Q_4$
and the enhancement of $r_4(Q_4)$ in the middle-$Q_4$ region [see Fig. 7(d) in
\cite{BaryRuZhang19}]. It is related to that the high-momentum pions are more
possibly emitted chaotically from excited states.

Table~\ref{Tab-r-intercept} presents the results of $r_3(Q_3)$ and $r_4(Q_4)$ at
$Q_3=10$~MeV/$c$ and $Q_4=30$~MeV/$c$, respectively, for the partially coherent pion emissions from a droplet with $p'=0.5$~GeV/$c$, $p'=0.7$~GeV/$c$, and $p'=\infty$.
One can see that the intercept results are almost consistent within the errors.

\vspace{0mm}
\begin{table}[hbt]
\begin{center}
\caption{Results of $r_3(Q_3\!=\!10$MeV$\!/\!c)$ and $r_4(Q_4\!=\!30$MeV$\!/\!c)$ for the partially coherent pion emissions of droplet with $p'\!=\!0.5$GeV$\!/\!c$, $p'\!=\!0.7$GeV$\!/\!c$, and $p'\!=\!\infty$.}
\footnotesize
\begin{tabular}{c|ccc}
\toprule
$p'$(GeV$\!/\!c)$&~0.5~&~0.7~&~$\infty$~\\
\hline
 & \multicolumn{3}{c}{$r_3(Q_3\!=\!10$MeV$\!/\!c)$} \\
\hline
$0.16\!<\!\!K_{T3}\!\!<\!0.3\,{\rm GeV}\!/c$~
&~$1.64\pm0.13$~&~~$1.51\pm0.10$~~&$~1.47\pm0.11$~\\
$0.3\!<\!K_{T3}\!<\!1\,{\rm GeV}\!/c$~
&~$1.67\pm0.14$~&~~$1.54\pm0.10$~~&~$1.45\pm0.12$~\\
\hline
 & \multicolumn{3}{c}{$r_4(Q_4\!=\!30$MeV$\!/\!c)$} \\
\hline
$~0.16\!<\!\!K_{T4}\!\!<\!0.3\,{\rm GeV}\!/c$~
&~$2.45\pm0.21$~&~~$2.29\pm0.18$~~&~$2.14\pm0.12$~\\
$0.3\!<\!K_{T4}\!<\!1\,{\rm GeV}\!/c$~
&~$2.07\pm0.24$~&~~$2.15\pm0.22$~~&~$1.98\pm0.19$~\\
\hline\hline
\end{tabular}
\label{Tab-r-intercept}
\end{center}
\end{table}

\section{Summary and discussion}
We investigate the three- and four-pion BECs in the granular source model with
coherent pion-emission droplets. The three- and four-pion correlation functions
and the normalized multi-pion correlation functions of the granular sources are
examined for completely coherent and momentum-dependent partially coherent
pion-emissions from a droplet. It is found that the intercepts of the multi-pion
correlation functions at the relative momenta near zero are sensitive to droplet
number in the granular source. They decrease with decreasing droplet number.

By comparing the three- and four-pion correlation functions of evolving granular
sources with the experimental data in Pb-Pb collisions at $\sqrt{s_{NN}}=2.76$ TeV
at the LHC, we find that the three-pion correlation functions for the evolving
granular sources with momentum-dependent partially coherent pion-emission droplets
are in basic agreement with the experimental data in the transverse-momentum
intervals $0.16<K_{T3}<0.3$~GeV/$c$ and $0.3<K_{T3}<1$~GeV/$c$. However, the model
results of four-pion correlation function are inconsistent with the experimental
data. To solve the puzzle of the multi-pion correlation functions, one needs more
consideration of the coherent pion emission.
For instance, the coherent emission may not only be dependent on
particle momentum magnitude but also be particle-azimuthal-angle dependent.
Figures.~\ref{EGs-C3-pp}(c) and \ref{EGs-C3-pp}(d) show that the model results in
the higher-transverse-momentum $K_{T4}$ interval have more enhancements compared
to the experimental data than those in the lower-transverse-momentum $K_{T4}$
intervals. This may indicate that the pions with small relative azimuthal angles,
thus larger $K_{T4}$, are possibly a coherent emission, though they have momenta
higher than the value of $p'$. The four-particle correlations are sensitive to the
coherent emission and should be investigated in more detail in model and experimental
data analyses. 

The normalized multi-pion correlation functions, defined as the ratios of the
multi-pion cumulant correlators to the two-pion correlator, can reduce the
influence of resonance decay on themselves. Our investigations indicate that
the normalized four-pion correlation function has an obvious enhancement in
$Q_4\sim\!100$-MeV$\!/c$ region for the wide transverse momentum interval
$0.3<K_{T4}<1$ GeV$\!/c$. It is related to that the high-momentum pions are
more possibly emitted chaotically from excited states.

Recently, D. Gangadharan proposed a technique of constructing three- and four-pion
correlation functions for partially coherent sources and estimating the source
coherence \cite{Gangadharan15}. Using this technique the ALICE Collaboration
analyzed the three- and four-pion correlation functions in Pb-Pb collisions at
$\sqrt{s_{NN}}=2.76$ TeV at the LHC \cite{ALICE-PRC16}. They find that the source
coherent fraction extracted from the four-pion correlation function cannot explain
the data of three-pion correlation function if the sources are assumed partially
coherent \cite{ALICE-PRC16}. How to consistently solve the suppressions of the
three- and four-pion BECs in a partially coherent source model is still an open
question.

Finally, it should be mentioned that the granular source model used in this
paper had the same model parameters as Ref. \cite{YangZhangRen17}, but with
the assumption of coherent pion emission from one droplet. We noted that this
assumption may increase the two-pion interferometry radii by $\sim 4\%$ on average
and decrease the two-pion chaoticity parameter $\lambda$ by $\sim 10\%$.
However, the assumption of coherent pion emission hardly changes the results of
the transverse-momentum spectrum and elliptic flow. Considering more realistic
pion coherent emission, using the Gangadharan's technique to granular source, and
investigating multi-pion BECs in more realistic model will be of interest.

\begin{acknowledgments}
This research was supported by the National Natural Science Foundation of China
under Grant Nos. 11675034 and 11275037.
\end{acknowledgments}

\vspace*{5mm}
\appendix
\section{Expression of four-pion correlation function of static granular source}
With the formulism developed in Refs. \cite{PraSieVis92,Zhang95}, one can get the
four-pion correlation function of the static granular source with completely chaotic
pion emission droplets as
\begin{widetext}
\begin{eqnarray}
&&C_4(\mk_1,\mk_2,\mk_3,\mk_4)=1+\frac{1}{n}\bigg[e^{-\mq_{12}^2 r_d^2} +e^{-\mq_{13}^2
r_d^2}+e^{-\mq_{14}^2 r_d^2}+e^{-\mq_{23}^2 r_d^2}+e^{-\mq_{24}^2 r_d^2} +e^{-\mq_{34}^2
r_d^2}\bigg] +\frac{(n\!-\!1)}{n}\bigg[e^{-\mq_{12}^2 (r_d^2+R_G^2)}\cr
&&\hspace*{40mm}
+\,e^{-\mq_{13}^2 (r_d^2+R_G^2)} +e^{-\mq_{14}^2 (r_d^2+R_G^2)}+e^{-\mq_{23}^2
(r_d^2+R_G^2)} +e^{-\mq_{24}^2 (r_d^2+R_G^2)} +e^{-\mq_{34}^2 (r_d^2+R_G^2)}
\bigg]\cr
&&\hspace*{34mm}
+\,\frac{2}{n^2}\bigg[e^{\!-\!(\mq_{12}^2+\mq_{13}^2+\mq_{23}^2)r_d^2/2}
\!+\!e^{\!-\!(\mq_{12}^2+\mq_{14}^2+\mq_{24}^2)r_d^2/2}\!+\!e^{\!-\!(\mq_{13}^2
+\mq_{14}^2+\mq_{34}^2)r_d^2/2}\!+\!e^{\!-\!(\mq_{23}^2+\mq_{24}^2+\mq_{34}^2)r_d^2/2}
\bigg] \cr
&&\hspace*{34mm}
+\frac{2(n-1)(n-2)}{n^2}\bigg[e^{-(\mq_{12}^2+\mq_{13}^2+\mq_{23}^2)(r_d^2+R_G^2)/2}
+e^{-(\mq_{12}^2+\mq_{14}^2+\mq_{24}^2)(r_d^2+R_G^2)/2}
\cr
&&\hspace*{40mm}
+\,e^{-(\mq_{13}^2+\mq_{14}^2+\mq_{34}^2)(r_d^2+R_G^2)/2} +e^{-(\mq_{23}^2+\mq_{24}^2+\mq_{34}^2)(r_d^2+R_G^2)/2}
\bigg] \cr
&&\hspace*{34mm}
+\frac{2(n-1)}{n^2}\bigg[e^{-(\mq_{12}^2+\mq_{13}^2+\mq_{23}^2)r_d^2/2} \Big(
e^{\!-\!\mq_{12}^2R_G^2}\!+\!e^{\!-\!\mq_{13}^2R_G^2}\!+\!e^{\!-\!\mq_{23}^2R_G^2}
\Big) +e^{-(\mq_{12}^2+\mq_{14}^2+\mq_{24}^2)r_d^2/2}\cr
&&\hspace*{40mm}
\times\,\Big(e^{\!-\!\mq_{12}^2 R_G^2} \!+\!e^{\!-\!\mq_{14}^2R_G^2}
\!+\!e^{\!-\!\mq_{24}^2 R_G^2}\Big) \!+\!e^{\!-\!(\mq_{13}^2+\mq_{14}^2
+\mq_{34}^2)r_d^2/2}\Big(e^{\!-\!\mq_{13}^2 R_G^2}+e^{\!-\!\mq_{14}^2
R_G^2}\!+\!e^{\!-\!\mq_{34}^2 R_G^2}\Big)
\cr
&&\hspace*{40mm}
+e^{-(\mq_{23}^2+\mq_{24}^2+\mq_{34}^2)r_d^2/2}\Big(e^{-\mq_{23}^2 R_G^2}
+e^{-\mq_{24}^2 R_G^2}+e^{-\mq_{34}^2 R_G^2}\Big)\bigg]
+\frac{1}{n^2}\bigg[e^{-(\mq_{12}^2+\mq_{34}^2)r_d^2}
\cr
&&\hspace*{40mm}
+\,e^{-(\mq_{13}^2+\mq_{24}^2)r_d^2} +e^{-(\mq_{14}^2+\mq_{23}^2)r_d^2} \bigg]
+\frac{(n-1)(n-2)(n-3)}{n^3}\bigg[e^{-(\mq_{12}^2+\mq_{34}^2)(r_d^2+R_G^2)}
\cr
&&\hspace*{40mm}
+\,e^{-(\mq_{13}^2+\mq_{24}^2)(r_d^2+R_G^2)} +e^{-(\mq_{14}^2
+\mq_{23}^2)(r_d^2+R_G^2)} \bigg]
+\frac{(n-1)}{n^2}\bigg[e^{-\mq_{12}^2 r_d^2} e^{-\mq_{34}^2 (r_d^2+R_G^2)}\cr
&&\hspace*{40mm}
+\,e^{\!-\!\mq_{13}^2 r_d^2} e^{\!-\!\mq_{24}^2 (r_d^2+\!R_G^2)}\!+\!e^{\!-\!\mq_{14}^2 r_d^2} e^{\!-\!\mq_{23}^2 (r_d^2+\!R_G^2)} \!+\!e^{\!-\!\mq_{23}^2 r_d^2}
e^{\!-\!\mq_{14}^2(r_d^2+\!R_G^2)} \!+\!e^{\!-\!\mq_{24}^2 r_d^2}
e^{\!-\!\mq_{13}^2(r_d^2+\!R_G^2)} \cr
&&\hspace*{40mm}
+\,e^{\!-\mq_{34}^2r_d^2}e^{\!-\mq_{12}^2(r_d^2+R_G^2)} \bigg] +\frac{(n-1)}{n^3}\bigg[e^{-(\mq_{12}^2+\mq_{34}^2)r_d^2} e^{-(\mq_{12}
+\mq_{34})^2 R_G^2} +e^{-(\mq_{12}^2+\mq_{34}^2)r_d^2} \cr
&&\hspace*{40mm}
\times\,e^{-(\mq_{12}
+\mq_{43})^2 R_G^2} +e^{-(\mq_{13}^2+\mq_{24}^2)r_d^2}
e^{\!-(\mq_{13}+\mq_{24})^2 R_G^2} +e^{\!-(\mq_{13}^2+\mq_{24}^2)r_d^2}
e^{\!-(\mq_{13}+\mq_{42})^2 R_G^2} \cr
&&\hspace*{40mm}
+\,e^{\!-(\mq_{14}^2+\mq_{23}^2)r_d^2} e^{\!-(\mq_{14}+\mq_{23})^2 R_G^2}
+e^{\!-(\mq_{14}^2+\mq_{23}^2)r_d^2} e^{\!-(\mq_{14}+\mq_{32})^2 R_G^2}\bigg]
\cr
&&\hspace*{34mm}
+\,\frac{2(n-1)(n-2)}{n^3}\bigg[e^{-(\mq_{12}^2+\mq_{34}^2)r_d^2} e^{-(\mq_{12}^2
+\mq_{34}^2) R_G^2/2} \Big(e^{-(\mq_{12}+\mq_{34})^2 R_G^2/2} +e^{-(\mq_{12}+
\mq_{43})^2 R_G^2/2}\Big)
\cr
&&\hspace*{40mm}
+\,e^{-(\mq_{13}^2+\mq_{24}^2)r_d^2}e^{-(\mq_{13}^2+\mq_{24}^2)R_G^2/2} \Big( e^{-(\mq_{13}+\mq_{24})^2R_G^2/2}+e^{-(\mq_{13}+\mq_{42})^2R_G^2/2} \Big)
 \cr
&&\hspace*{40mm}
+\,e^{-(\mq_{14}^2+\mq_{23}^2)r_d^2} e^{-(\mq_{14}^2+\mq_{23}^2)R_G^2/2} \Big(
e^{-(\mq_{14}+\mq_{23})^2R_G^2/2}+e^{-(\mq_{14}+\mq_{32})^2R_G^2/2} \Big) \bigg]
\cr
&&\hspace*{34mm}
+\,\frac{2}{n^3}\bigg[e^{-(\mq_{12}^2+\mq_{23}^2+\mq_{34}^2+\mq_{41}^2)r_d^2/2}
+e^{-(\mq_{12}^2+\mq_{24}^2+\mq_{43}^2+\mq_{31}^2)r_d^2/2}
+e^{-(\mq_{13}^2+\mq_{32}^2+\mq_{24}^2+\mq_{41}^2)r_d^2/2}\bigg]
\cr
&&\hspace*{34mm}
+\,\frac{2(n\!-\!1)(n\!-\!2)(n\!-\!3)}{n^3}\bigg[e^{\!-\!(\mq_{12}^2+\mq_{23}^2
+\mq_{34}^2+\mq_{41}^2)(r_d^2+\!R_G^2)/2} +e^{\!-\!(\mq_{12}^2+\mq_{24}^2
+\mq_{43}^2+\mq_{31}^2)(r_d^2+\!R_G^2)/2}
\cr
&&\hspace*{40mm}
+\,e^{\!-\!(\mq_{13}^2+\mq_{32}^2+\mq_{24}^2+\mq_{41}^2)(r_d^2+\!R_G^2)/2}\bigg]
+\frac{2(n\!-\!1)}{n^3}\bigg[e^{\!-\!(\mq_{12}^2+\mq_{23}^2+\mq_{34}^2+\mq_{41}^2)
r_d^2/2}\Big(e^{-\mq_{12}^2R_G^2}
\cr
&&\hspace*{40mm}
+\,e^{-\mq_{23}^2R_G^2}\!+\!e^{-\mq_{34}^2R_G^2}+e^{-\mq_{41}^2R_G^2}\Big)
\!+\!e^{-(\mq_{12}^2+\mq_{24}^2+\mq_{43}^2\!+\!\mq_{31}^2)r_d^2/2}\Big(e^{-\mq_{12}^2
R_G^2}\!+\!e^{-\mq_{24}^2R_G^2}
\cr
&&\hspace*{40mm}
+\,e^{\!-\!\mq_{43}^2R_G^2}\!+\!e^{\!-\!\mq_{31}^2R_G^2}\Big)\!+\!e^{\!-\!(\mq_{13}^2
+\mq_{32}^2+\mq_{24}^2+\mq_{41}^2)r_d^2/2} \Big(e^{\!-\!\mq_{13}^2R_G^2}
\!+\!e^{\!-\!\mq_{32}^2R_G^2}\!+\!e^{\!-\!\mq_{24}^2R_G^2}\!+\!e^{\!-\!\mq_{41}^2R_G^2}
\Big)\bigg] \cr
&&\hspace*{34mm}
+\,\frac{(n-1)}{n^3}\bigg[e^{-(\mq_{12}^2+\mq_{23}^2+\mq_{34}^2+\mq_{41}^2)r_d^2/2}
\Big(e^{-(\mq_{12}+\mq_{23})^2R_G^2}+e^{-(\mq_{12}+\mq_{34})^2R_G^2}
+e^{-(\mq_{12}+\mq_{41})^2R_G^2}
\cr
&&\hspace*{40mm}
+\,e^{-(\mq_{23}+\mq_{34})^2R_G^2}+e^{-(\mq_{23}+\mq_{41})^2R_G^2}
+e^{-(\mq_{34}+\mq_{41})^2R_G^2}\Big)
+e^{-(\mq_{12}^2+\mq_{24}^2+\mq_{43}^2+\mq_{31}^2)r_d^2/2}
\cr
&&\hspace*{40mm}
\times\,\Big(e^{\!-\!(\mq_{12}+\mq_{24})^2\!R_G^2}\!+\!e^{\!-\!(\mq_{12}+\mq_{43})^2
\!R_G^2}\!+\!e^{\!-\!(\mq_{12}+\mq_{31})^2\!R_G^2}\!+\!e^{\!-\!(\mq_{24}+\mq_{43})^2
\!R_G^2}\!+\!e^{\!-\!(\mq_{24}+\mq_{31})^2\!R_G^2}
\cr
&&\hspace*{40mm}
+\,e^{-(\mq_{43}+\mq_{31})^2R_G^2}\Big) +e^{-(\mq_{13}^2+\mq_{32}^2+\mq_{24}^2
+\mq_{41}^2)r_d^2/2} \Big(e^{-(\mq_{13}+\mq_{32})^2R_G^2}+e^{-(\mq_{13}+\mq_{24})^2
R_G^2} \cr
&&\hspace*{40mm}
+e^{-(\mq_{13}+\mq_{41})^2R_G^2}+e^{-(\mq_{32}+\mq_{24})^2R_G^2}
+e^{-(\mq_{32}+\mq_{41})^2R_G^2}+e^{-(\mq_{24}+\mq_{41})^2R_G^2} \Big)
\bigg] \cr
&&\hspace*{34mm}
+\,\frac{2(n-1)(n-2)}{n^3}\bigg[e^{-(\mq_{12}^2+\mq_{23}^2+\mq_{34}^2+\mq_{41}^2)
r_d^2/2} \Big(e^{-(\mq_{12}^2+\mq_{13}^2+\mq_{23}^2)R_G^2/2}
\!+\!\,e^{-(\mq_{12}^2 +\mq_{14}^2+\mq_{24}^2)R_G^2/2}
\cr
&&\hspace*{40mm}
+\,e^{-(\mq_{13}^2 +\mq_{14}^2+\mq_{34}^2)R_G^2/2}
+e^{-(\mq_{23}^2 +\mq_{24}^2+\mq_{34}^2)R_G^2/2}
+e^{-(\mq_{12}^2+\mq_{34}^2)R_G^2/2} e^{-(\mq_{12}+\mq_{43})^2R_G^2/2}
\cr
&&\hspace*{40mm}
+\,e^{-(\mq_{14}^2+\mq_{23}^2)R_G^2/2} e^{-(\mq_{14}+\mq_{32})^2R_G^2/2}\Big)
\!+\!e^{-(\mq_{12}^2+\mq_{24}^2+\mq_{43}^2+\mq_{31}^2)r_d^2/2} \Big(
e^{-(\mq_{12}^2+\mq_{13}^2+\mq_{23}^2)R_G^2/2}
\cr
&&\hspace*{40mm}
+\,e^{-(\mq_{12}^2 +\mq_{14}^2+\mq_{24}^2)R_G^2/2}
+e^{-(\mq_{13}^2 +\mq_{14}^2+\mq_{34}^2)R_G^2/2}
+e^{-(\mq_{23}^2 +\mq_{24}^2+\mq_{24}^2)R_G^2/2}
\cr
&&\hspace*{40mm}
+\,e^{-(\mq_{13}^2+\mq_{24}^2)R_G^2/2} e^{-(\mq_{13}+\mq_{42})^2R_G^2/2}
+e^{-(\mq_{12}^2+\mq_{34}^2)R_G^2/2} e^{-(\mq_{12}+\mq_{43})^2R_G^2/2}\Big)
\cr
&&\hspace*{40mm}
+\,e^{-(\mq_{13}^2 +\mq_{32}^2+\mq_{24}^2+\mq_{41}^2)r_d^2/2}
\Big(e^{-(\mq_{12}^2+\mq_{13}^2+\mq_{23}^2)R_G^2/2} +e^{-(\mq_{12}^2
+\mq_{14}^2+\mq_{24}^2)R_G^2/2}
\cr
&&\hspace*{40mm}
+\,e^{-(\mq_{13}^2 +\mq_{14}^2+\mq_{34}^2)R_G^2/2}
+e^{-(\mq_{23}^2 +\mq_{24}^2+\mq_{34}^2)R_G^2/2}
+e^{-(\mq_{13}^2+\mq_{24}^2)R_G^2/2} e^{-(\mq_{13}+\mq_{42})^2R_G^2/2}
\cr
&&\hspace*{40mm}
+\,e^{-(\mq_{14}^2+\mq_{23}^2)R_G^2/2} e^{-(\mq_{14}+\mq_{32})^2R_G^2/2}\Big)\bigg],
\label{GsC4A1}
\end{eqnarray}
\end{widetext}
where $\mq_{ij}=\mk_i-\mk_j~(i,j=1,2,3,4)$ and $n$ is the droplet number in the
granular source. In Eq.~(\ref{GsC4A1}), the first and second square brackets are the
correlations of two pions which are emitted from one droplet and from two different
droplets, respectively; the third and forth square brackets are the pure triplet correlations of three pions which are emitted from one droplet and from three different
droplets, respectively; the fifth square brackets are the pure triplet correlations of three pions where two pions are emitted from one droplet and another pion is emitted
from a different droplet; the sixth and seventh square brackets are the correlations
of double pion pairs where each pair is emitted from one droplet and the four pions
are emitted from four different droplets, respectively; the eighth square brackets are
the correlations of double pion pairs where one pion pair is emitted from one droplet and the two pions of another pair are emitted from two different droplets; the ninth square brackets are the correlations of double pion pairs where the two pions of a pair are emitted from two different droplets and the two pions of another pair are emitted respectively from the two droplets also; the tenth square brackets are the correlations of double pion pairs where the two pions of a pair are emitted from two different droplets and the two pions of another pair are emitted respectively from one of the
same droplets and from another droplet; the eleventh and twelfth square brackets are the pure quadruplet correlations of four pions emitted from one droplet and from four different droplets, respectively; the thirteenth square brackets are the pure quadruplet correlations of four pions where three pions are emitted from one droplet and another
pion is emitted from a different droplet; the fourteenth square brackets are the pure quadruplet correlations of four pions where two pions are emitted from one
droplet and the other two pions are emitted from another droplet; and finally, the
fifteenth square brackets are the pure quadruplet correlations of four pions where two
pions are emitted from one droplet and the other two pions are respectively emitted from two different droplets.

The four-pion correlation function of granular source is complex which includes the
relative angles of two relative momenta in the double pair correlations and pure
quadruplet correlations. For the completely chaotic pion emission droplet,
$C_4(\mk_1,\mk_2,\mk_3,\mk_4)=24$ when $\mq_{ij}=0~(i,j=1,2,3,4)$. For a partially
coherent pion-emission from a droplet, if the coherent emission in a droplet have the
same Gaussian distribution of the chaotic pion emission and with a constant ratio
of coherent emission contribution $b_c$ to chaotic emission contribution $b_\chi$,
$\gamma=b_c/b_\chi$, the terms of two, pure triplet, double pair, and pure quadruplet
pion correlations will reduce by the factors \cite{Liu86,LiuPhD}, $\lambda=(1+2\gamma)
/(1+\gamma)^2$, $\xi=(1+3\gamma)/(1+\gamma)^3$, $\lambda^2$, and $\eta=(1+4\gamma)
/(1+\gamma)^4$, respectively. In this case, the four-pion correlation function of the
granular source with partially coherent pion-emission droplets is given by,
\begin{widetext}
\begin{eqnarray}
&&C_4(\mk_1,\mk_2,\mk_3,\mk_4)=1+\frac{\lambda}{n}\bigg[e^{-\mq_{12}^2 r_d^2}
+e^{-\mq_{13}^2r_d^2}+e^{-\mq_{14}^2 r_d^2}+e^{-\mq_{23}^2 r_d^2}
+e^{-\mq_{24}^2 r_d^2} +e^{-\mq_{34}^2r_d^2}\bigg]
+\frac{(n\!-\!1)}{n}\bigg[e^{-\mq_{12}^2 (r_d^2+R_G^2)}
\cr
&&\hspace*{40mm}
+\,e^{-\mq_{13}^2 (r_d^2+R_G^2)} +e^{-\mq_{14}^2 (r_d^2+R_G^2)}+e^{-\mq_{23}^2
(r_d^2+R_G^2)} +e^{-\mq_{24}^2 (r_d^2+R_G^2)} +e^{-\mq_{34}^2 (r_d^2+R_G^2)}\bigg]
\cr
&&\hspace*{34mm}
+\,\frac{2\xi}{n^2}\bigg[e^{\!-\!(\mq_{12}^2+\mq_{13}^2+\mq_{23}^2)r_d^2/2}
\!+\!e^{\!-\!(\mq_{12}^2+\mq_{14}^2+\mq_{24}^2)r_d^2/2}\!+\!e^{\!-\!(\mq_{13}^2
+\mq_{14}^2+\mq_{34}^2)r_d^2/2}\!+\!e^{\!-\!(\mq_{23}^2+\mq_{24}^2+\mq_{34}^2)r_d^2/2}
\bigg] \cr
&&\hspace*{34mm}
+\frac{2(n-1)(n-2)}{n^2}\bigg[e^{-(\mq_{12}^2+\mq_{13}^2+\mq_{23}^2)(r_d^2+R_G^2)/2}
+e^{-(\mq_{12}^2+\mq_{14}^2+\mq_{24}^2)(r_d^2+R_G^2)/2}
\cr
&&\hspace*{40mm}
+\,e^{-(\mq_{13}^2+\mq_{14}^2+\mq_{34}^2)(r_d^2+R_G^2)/2} +e^{-(\mq_{23}^2+\mq_{24}^2+\mq_{34}^2)(r_d^2+R_G^2)/2} \bigg]
\cr
&&\hspace*{34mm}
+\frac{2(n-1)\lambda}{n^2}\bigg[e^{-(\mq_{12}^2+\mq_{13}^2+\mq_{23}^2)r_d^2/2} \Big(
e^{\!-\!\mq_{12}^2R_G^2}\!+\!e^{\!-\!\mq_{13}^2R_G^2}\!+\!e^{\!-\!\mq_{23}^2R_G^2}
\Big) +e^{-(\mq_{12}^2+\mq_{14}^2+\mq_{24}^2)r_d^2/2}
\cr
&&\hspace*{40mm}
\times\,\Big(e^{\!-\!\mq_{12}^2 R_G^2} \!+\!e^{\!-\!\mq_{14}^2R_G^2}
\!+\!e^{\!-\!\mq_{24}^2 R_G^2}\Big) \!+\!e^{\!-\!(\mq_{13}^2+\mq_{14}^2
+\mq_{34}^2)r_d^2/2}\Big(e^{\!-\!\mq_{13}^2 R_G^2}+e^{\!-\!\mq_{14}^2
R_G^2}\!+\!e^{\!-\!\mq_{34}^2 R_G^2}\Big)
\cr
&&\hspace*{40mm}
+e^{-(\mq_{23}^2+\mq_{24}^2+\mq_{34}^2)r_d^2/2}\Big(e^{-\mq_{23}^2 R_G^2}
+e^{-\mq_{24}^2 R_G^2}+e^{-\mq_{34}^2 R_G^2}\Big)\bigg]
+\frac{\lambda^2}{n^2}\bigg[e^{-(\mq_{12}^2+\mq_{34}^2)r_d^2}
\cr
&&\hspace*{40mm}
+\,e^{-(\mq_{13}^2+\mq_{24}^2)r_d^2} +e^{-(\mq_{14}^2+\mq_{23}^2)r_d^2} \bigg]
+\frac{(n-1)(n-2)(n-3)}{n^3}\bigg[e^{-(\mq_{12}^2+\mq_{34}^2)(r_d^2+R_G^2)}
\cr
&&\hspace*{40mm}
+\,e^{-(\mq_{13}^2+\mq_{24}^2)(r_d^2+R_G^2)} +e^{-(\mq_{14}^2
+\mq_{23}^2)(r_d^2+R_G^2)} \bigg]
+\frac{(n-1)\lambda}{n^2}\bigg[e^{-\mq_{12}^2 r_d^2} e^{-\mq_{34}^2 (r_d^2+R_G^2)}
\cr
&&\hspace*{40mm}
+\,e^{\!-\!\mq_{13}^2 r_d^2} e^{\!-\!\mq_{24}^2 (r_d^2+\!R_G^2)}\!+\!e^{\!-\!\mq_{14}^2 r_d^2} e^{\!-\!\mq_{23}^2 (r_d^2+\!R_G^2)} \!+\!e^{\!-\!\mq_{23}^2 r_d^2}
e^{\!-\!\mq_{14}^2(r_d^2+\!R_G^2)} \!+\!e^{\!-\!\mq_{24}^2 r_d^2}
e^{\!-\!\mq_{13}^2(r_d^2+\!R_G^2)}
\cr
&&\hspace*{40mm}
+\,e^{\!-\mq_{34}^2r_d^2}e^{\!-\mq_{12}^2(r_d^2+R_G^2)} \bigg] +\frac{(n-1)}{n^3}\bigg[e^{-(\mq_{12}^2+\mq_{34}^2)r_d^2} e^{-(\mq_{12}
+\mq_{34})^2 R_G^2} +e^{-(\mq_{12}^2+\mq_{34}^2)r_d^2}
\cr
&&\hspace*{40mm}
\times\,e^{-(\mq_{12}+\mq_{43})^2 R_G^2} +e^{-(\mq_{13}^2+\mq_{24}^2)r_d^2}
e^{\!-(\mq_{13}+\mq_{24})^2 R_G^2} +e^{\!-(\mq_{13}^2+\mq_{24}^2)r_d^2}
e^{\!-(\mq_{13}+\mq_{42})^2 R_G^2}
\cr
&&\hspace*{40mm}
+\,e^{\!-(\mq_{14}^2+\mq_{23}^2)r_d^2} e^{\!-(\mq_{14}+\mq_{23})^2 R_G^2}
+e^{\!-(\mq_{14}^2+\mq_{23}^2)r_d^2} e^{\!-(\mq_{14}+\mq_{32})^2 R_G^2}\bigg]
\cr
&&\hspace*{34mm}
+\,\frac{2(n-1)(n-2)}{n^3}\bigg[e^{-(\mq_{12}^2+\mq_{34}^2)r_d^2} e^{-(\mq_{12}^2
+\mq_{34}^2) R_G^2/2} \Big(e^{-(\mq_{12}+\mq_{34})^2 R_G^2/2} +e^{-(\mq_{12}+
\mq_{43})^2 R_G^2/2}\Big)
\cr
&&\hspace*{40mm}
+\,e^{-(\mq_{13}^2+\mq_{24}^2)r_d^2}e^{-(\mq_{13}^2+\mq_{24}^2)R_G^2/2} \Big( e^{-(\mq_{13}+\mq_{24})^2R_G^2/2}+e^{-(\mq_{13}+\mq_{42})^2R_G^2/2} \Big)
\cr
&&\hspace*{40mm}
+\,e^{-(\mq_{14}^2+\mq_{23}^2)r_d^2} e^{-(\mq_{14}^2+\mq_{23}^2)R_G^2/2} \Big(
e^{-(\mq_{14}+\mq_{23})^2R_G^2/2}+e^{-(\mq_{14}+\mq_{32})^2R_G^2/2} \Big) \bigg]
\cr
&&\hspace*{34mm}
+\,\frac{2\eta}{n^3}\bigg[e^{-(\mq_{12}^2+\mq_{23}^2+\mq_{34}^2+\mq_{41}^2)r_d^2/2}
+e^{-(\mq_{12}^2+\mq_{24}^2+\mq_{43}^2+\mq_{31}^2)r_d^2/2}
+e^{-(\mq_{13}^2+\mq_{32}^2+\mq_{24}^2+\mq_{41}^2)r_d^2/2}\bigg]
\cr
&&\hspace*{34mm}
+\,\frac{2(n\!-\!1)(n\!-\!2)(n\!-\!3)}{n^3}\bigg[e^{\!-\!(\mq_{12}^2+\mq_{23}^2
+\mq_{34}^2+\mq_{41}^2)(r_d^2+\!R_G^2)/2} +e^{\!-\!(\mq_{12}^2+\mq_{24}^2
+\mq_{43}^2+\mq_{31}^2)(r_d^2+\!R_G^2)/2}
\cr
&&\hspace*{40mm}
+\,e^{\!-\!(\mq_{13}^2+\mq_{32}^2+\mq_{24}^2+\mq_{41}^2)(r_d^2+\!R_G^2)/2}\bigg]
+\frac{2(n\!-\!1)\xi}{n^3}\bigg[e^{\!-\!(\mq_{12}^2+\mq_{23}^2+\mq_{34}^2+\mq_{41}^2)
r_d^2/2}\Big(e^{-\mq_{12}^2R_G^2}
\cr
&&\hspace*{40mm}
+\,e^{-\mq_{23}^2R_G^2}\!+\!e^{-\mq_{34}^2R_G^2}+e^{-\mq_{41}^2R_G^2}\Big)
\!+\!e^{-(\mq_{12}^2+\mq_{24}^2+\mq_{43}^2\!+\!\mq_{31}^2)r_d^2/2}\Big(e^{-\mq_{12}^2
R_G^2}\!+\!e^{-\mq_{24}^2R_G^2}
\cr
&&\hspace*{40mm}
+\,e^{\!-\!\mq_{43}^2R_G^2}\!+\!e^{\!-\!\mq_{31}^2R_G^2}\Big)\!+\!e^{\!-\!(\mq_{13}^2
+\mq_{32}^2+\mq_{24}^2+\mq_{41}^2)r_d^2/2} \Big(e^{\!-\!\mq_{13}^2R_G^2}
\!+\!e^{\!-\!\mq_{32}^2R_G^2}\!+\!e^{\!-\!\mq_{24}^2R_G^2}\!+\!e^{\!-\!\mq_{41}^2R_G^2}
\Big)\bigg]
\cr
&&\hspace*{34mm}
+\,\frac{(n-1)\lambda^2}{n^3}\bigg[e^{-(\mq_{12}^2+\mq_{23}^2+\mq_{34}^2+\mq_{41}^2)
r_d^2/2} \Big(e^{-(\mq_{12}+\mq_{23})^2R_G^2}+e^{-(\mq_{12}+\mq_{34})^2R_G^2}
+e^{-(\mq_{12}+\mq_{41})^2R_G^2}
\cr
&&\hspace*{40mm}
+\,e^{-(\mq_{23}+\mq_{34})^2R_G^2}+e^{-(\mq_{23}+\mq_{41})^2R_G^2}
+e^{-(\mq_{34}+\mq_{41})^2R_G^2}\Big)
+e^{-(\mq_{12}^2+\mq_{24}^2+\mq_{43}^2+\mq_{31}^2)r_d^2/2}
\cr
&&\hspace*{40mm}
\times\,\Big(e^{\!-\!(\mq_{12}+\mq_{24})^2\!R_G^2}\!+\!e^{\!-\!(\mq_{12}+\mq_{43})^2
\!R_G^2}\!+\!e^{\!-\!(\mq_{12}+\mq_{31})^2\!R_G^2}\!+\!e^{\!-\!(\mq_{24}+\mq_{43})^2
\!R_G^2}\!+\!e^{\!-\!(\mq_{24}+\mq_{31})^2\!R_G^2}
\cr
&&\hspace*{40mm}
+\,e^{-(\mq_{43}+\mq_{31})^2R_G^2}\Big) +e^{-(\mq_{13}^2+\mq_{32}^2+\mq_{24}^2
+\mq_{41}^2)r_d^2/2} \Big(e^{-(\mq_{13}+\mq_{32})^2R_G^2}+e^{-(\mq_{13}+\mq_{24})^2
R_G^2}
\cr
&&\hspace*{40mm}
+e^{-(\mq_{13}+\mq_{41})^2R_G^2}+e^{-(\mq_{32}+\mq_{24})^2R_G^2}
+e^{-(\mq_{32}+\mq_{41})^2R_G^2}+e^{-(\mq_{24}+\mq_{41})^2R_G^2} \Big) \bigg]
\cr
&&\hspace*{34mm}
+\,\frac{2(n-1)(n-2)\lambda}{n^3}\bigg[e^{-(\mq_{12}^2+\mq_{23}^2+\mq_{34}^2
+\mq_{41}^2)r_d^2/2} \Big(e^{-(\mq_{12}^2+\mq_{13}^2+\mq_{23}^2)R_G^2/2}
\!+\!\,e^{-(\mq_{12}^2 +\mq_{14}^2+\mq_{24}^2)R_G^2/2}
\cr
&&\hspace*{40mm}
+\,e^{-(\mq_{13}^2 +\mq_{14}^2+\mq_{34}^2)R_G^2/2}
+e^{-(\mq_{23}^2 +\mq_{24}^2+\mq_{34}^2)R_G^2/2}
+e^{-(\mq_{12}^2+\mq_{34}^2)R_G^2/2} e^{-(\mq_{12}+\mq_{43})^2R_G^2/2}
\cr
&&\hspace*{40mm}
+\,e^{-(\mq_{14}^2+\mq_{23}^2)R_G^2/2} e^{-(\mq_{14}+\mq_{32})^2R_G^2/2}\Big)
\!+\!e^{-(\mq_{12}^2+\mq_{24}^2+\mq_{43}^2+\mq_{31}^2)r_d^2/2} \Big(
e^{-(\mq_{12}^2+\mq_{13}^2+\mq_{23}^2)R_G^2/2}
\cr
&&\hspace*{40mm}
+\,e^{-(\mq_{12}^2 +\mq_{14}^2+\mq_{24}^2)R_G^2/2}
+e^{-(\mq_{13}^2 +\mq_{14}^2+\mq_{34}^2)R_G^2/2}
+e^{-(\mq_{23}^2 +\mq_{24}^2+\mq_{24}^2)R_G^2/2}
\cr
&&\hspace*{40mm}
+\,e^{-(\mq_{13}^2+\mq_{24}^2)R_G^2/2} e^{-(\mq_{13}+\mq_{42})^2R_G^2/2}
+e^{-(\mq_{12}^2+\mq_{34}^2)R_G^2/2} e^{-(\mq_{12}+\mq_{43})^2R_G^2/2}\Big)
\cr
&&\hspace*{40mm}
+\,e^{-(\mq_{13}^2 +\mq_{32}^2+\mq_{24}^2+\mq_{41}^2)r_d^2/2}
\Big(e^{-(\mq_{12}^2+\mq_{13}^2+\mq_{23}^2)R_G^2/2} +e^{-(\mq_{12}^2
+\mq_{14}^2+\mq_{24}^2)R_G^2/2}
\cr
&&\hspace*{40mm}
+\,e^{-(\mq_{13}^2 +\mq_{14}^2+\mq_{34}^2)R_G^2/2}
+e^{-(\mq_{23}^2 +\mq_{24}^2+\mq_{34}^2)R_G^2/2}
+e^{-(\mq_{13}^2+\mq_{24}^2)R_G^2/2} e^{-(\mq_{13}+\mq_{42})^2R_G^2/2}
\cr
&&\hspace*{40mm}
+\,e^{-(\mq_{14}^2+\mq_{23}^2)R_G^2/2} e^{-(\mq_{14}+\mq_{32})^2R_G^2/2}\Big)\bigg],
\label{GsC4A2}
\end{eqnarray}
\end{widetext}

For the granular source with completely coherent pion-emission droplets, the factors
$\lambda$, $\xi$, and $\eta$ are zero, and the four-pion correlation function reduces
to a simple formula [see Eq.~(\ref{GsC4})]. In this case, the maximum of $C_4(\mk_1,
\mk_2,\mk_3,\mk_4)$ at $\mq_{ij}=0~(i,j=1,2,3,4)$ is $n$-dependent.

\end{document}